\documentclass[12pt,a4papper]{article}

\usepackage{amssymb}

\begin{document}

\title{The Pioneer riddle, the quantum vacuum and the acceleration of
light\large{\thanks{Shortened title: The Pioneer
riddle and the light velocity}}}
\author{Antonio F. Ra\~nada\\Departamento de F\'{\i}sica Aplicada III,
Universidad Complutense, \\28040 Madrid, Spain \thanks{E-mail
address: afr@fis.ucm.es}}

\date{May 26, 2003}

\maketitle

\begin{abstract}
It is shown that the same phenomenological Newtonian model
recently proposed, which accounts for the cosmological evolution
of the fine structure constant, suggests furthermore an
explanation of the unmodelled acceleration $a_{\rm P}\simeq
8.5\times 10^{-10}\mbox{ m/s}^2$  of the Pioneer 10/11 spaceships
reported by Anderson {\em et al} in 1998. In the view presented
here,  the permittivity and permeability of empty space are
decreasing adiabatically, and the light is accelerating therefore,
as a consequence of the progressive attenuation of the quantum
vacuum due to the combined effect of its gravitational interaction
with all the expanding universe and the fourth Heisenberg
relation. It is argued that the spaceships might not have any
extra acceleration (but would follow instead the unchanged Newton
laws), the observed effect being due to an adiabatic acceleration
of the light equal to $a_{\rm P}$, which has the same
observational radio signature as the anomalous acceleration of the
Pioneers.

\end{abstract}


{\bf Introduction and purpose.} In a previous paper \cite{Ran02},
an explanation was proposed for the cosmological variation of the
fine structure constant observed by Webb {\em et al} \cite{Web01},
which is based in the gravitational interaction of the quantum
vacuum with all the universe. As was argued there, the quantum
vacuum must thin or attenuate adiabatically  along the history of
the universe with the consequent decrease of its permittivity and
its permeability. This causes, in turn,  a time dependent increase
of the fine structure constant, which agrees well with the
observations by Webb {\em et al}. The reader is referred to
\cite{Ran02} for all the necessary details. This letter shows that
the same model offers besides an explanation for the anomalous
Pioneer's acceleration.

{\bf The anomalous Pioneer's acceleration.} A tiny but significant
anomaly in the motion of the Pioneer 10/11 spaceships was reported
by Anderson {\em et al} \cite{And98} in 1998: the solar attraction
seems to be slightly larger than what predicted by Newton's laws.
The Nasa analysis of the data from the two missions showed in the
motion of both spacecrafts  an extra unmodelled constant
acceleration towards the Sun, equal to $a_{\rm P}\simeq 8.5\times
10^{-10}$ m/s$^2$. The data from the Galileo and Ulysses
spaceships confirmed the effect. Surprisingly, no similar extra
acceleration was found in the case of the planets, as it would be
required by the equivalence principle if the effect were due to
gravitational forces.  Anderson {\em et al} concluded ``it is
interesting to speculate on the possibility that the origin of the
anomalous signal is new physics." In spite of a thorough search,
no reason could be found as yet for that extra acceleration (see
\cite{And02} for a detailed review of the problem and of the
observational techniques involved).

In the explanation suggested here, there is indeed a genuine extra
blue shift of the radio waves from the Pioneers ({\em i.e.} the
Nasa team observed a real existing effect). The spaceships,
however, followed the exact trajectories predicted by Newton
unchanged law of Gravitation, without any extra pull from the Sun,
the observed effect being not due to any kind of unknown
acceleration of the ships but to an acceleration of the light.
Indeed, the model proposed in \cite{Ran02} to explain the
cosmological evolution of the fine structure constant predicts an
adiabatic acceleration of light which, at present time, would be
of the same order as $H_0c= 6.9\times 10^{-10}\mbox{ m/s}^2$
($H_0$ being the Hubble parameter), if two coefficients related to
the renormalization effects of the quantum vacuum are of order
one. Such acceleration would be due to an adiabatic decrease of
the permittivity and the permeability of empty space, consequence
of the decrease of the quantum vacuum density, produced by the
combined effects of the fourth Heisenberg relation and the
universe expansion on the quantum vacuum. It will be shown
moreover that an adiabatic acceleration of the light has the same
observational signature as a blue shift of the radio waves due to
an acceleration $a_{\rm P}$ of the Pioneers towards the Sun.

{\bf Summary of the model.} The model used in \cite{Ran02}  is
based on the effect of the gravitational potential $\Phi$ on the
density of the quantum vacuum, which is treated phenomenologically
as a transparent optical medium (note that as $\Phi$ is the
potential due to all the universe, this model is close in spirit
to the Mach principle). As the virtual particles in the vacuum
have a gravitational potential energy $E\Phi /c^2$, $E$ being the
non-gravitational energy, the fourth Heisenberg relation implies
that their average lifetime depends on $\Phi$, and consequently
the very density of the vacuum as well. More precisely, their
average lifetime in a gravitational potential is $\tau _\Phi =\tau
_0/(1+\Phi /c^2)$, $\tau _0$ being its value with $\Phi =0$. As
shown in \cite{Ran02}, a consequence is that, since the relative
permittivity and permeability of empty space must depend on the
gravitational potential $\Phi ({\bf r},t)$, they vary in
spacetime,  their expressions at first order being
\begin{equation}
\epsilon _{\rm r}({\bf r},t)=1-\beta [\Phi ({\bf r},t) -\Phi _{\rm
E} ]/c^2,\;\;\;\; \mu _{\rm r}({\bf r},t)=1-\gamma [\Phi ({\bf
r},t)-\Phi _{\rm E} ]/c^2, \label{10}
\end{equation}
where $\Phi _{\rm E}$ is the gravitational potential today and at
a reference terrestrial laboratory, and $\beta$ and $\gamma$ are
certain coefficients, which must be positive since the quantum
vacuum is dielectric but paramagnetic (its effect on the magnetic
field is due to the magnetic moments of the virtual pairs). This
implies that, at first order, the light velocity at a generic
spacetime point must be equal to
\begin{equation}
c({\bf r},t) = c[1+(\beta +\gamma )(\Phi ({\bf r},t) -\Phi _{\rm
E} )/(2c^2)], \label{15}
\end{equation}
$c$ being its value now at Earth (the constant in the tables),
with corresponding variations for the observed electron charge and
the fine structure constant. It follows from (\ref{15}) that
$c({\bf r},t)$ is smaller where $\Phi$ is more negative (or less
positive), {\it i.e.} it decreases when approaching massive
objects, but increases monotonously in time since the galaxies are
separating because of the expansion. Note that the two kinds of
variation of $\Phi ({\bf r},t)$ due  to the changes of ${\bf r}$
and $t$, respectively, have different effects. The first causes
the light to behave as in an ordinary optical medium, in such a
way that the frequency is constant during the propagation, while
the wavelength and the light velocity change according to the
value of a refractive index as in an inhomogeneous transparent
optical medium. It will be shown here that the second causes an
adiabatic increase of the light velocity and of the frequency, the
wavelength remaining constant.   The former is describable with a
refractive index $n({\bf r},t)\; [=\{1+(\beta +\gamma )(\Phi ({\bf
r},t) -\Phi _{\rm E} )/(2c^2)\}^{-1}]$. The latter arises because
the expansion implies, as noted before, that $\Phi$ is increasing
and $\epsilon _{\rm r}$, $\mu _{\rm r}$ are decreasing, with the
corresponding acceleration of light. It must be emphasized that
while the former is either positive or negative, according to how
much matter is around, the latter is secular and consists in a
monotonous adiabatic increase in the light velocity and the
frequency, as is shown in the following, which is however
negligible in terrestrial laboratory experiments. This is
important since one or the other of the two variations can be
neglected in some interesting cases.

A variation of $c$ should not be a matter of concern. Einstein
himself made the following enlightening comment in 1912 ``the
constancy of the velocity of light can be maintained only insofar
as one restricts oneself to spatio-temporal regions of constant
gravitational potential. This is where, in my opinion, the limit
of validity of the principle of the constancy of the velocity of
light---though not of the principle of relativity---and therewith
the limit of validity of our current theory of relativity lies"
\cite{Ein12}. Note that Einstein states clearly (i) that a
variation of $c$ does not imply necessarily a violation of the
principle of relativity if $c$ depends on the potential, as it
happens in eq. (\ref{15}) where $c({\bf r},t)=c[\Phi ({\bf
r},t)]$, and (ii) that we must distinguish carefully between the
principle of relativity and any particular theory of relativity.

{\bf  The attenuation of the quantum vacuum and the time variation
of the light velocity.} Note that $\Phi ({\bf r},t)$ and $\Phi
_{\rm E}$ are the sum of the space averaged potential of all the
universe $\Phi _{\rm av}(t)$ plus the contributions of the nearby
inhomogeneities $\Phi _{\rm inh}$, in the case of a terrestrial
laboratory the Earth, the Sun and the Milky Way. The former is
time dependent because of the general expansion, while the latter
is constant at Earth since these three objects are not expanding.
This means that at the Earth surface, the effect of the
inhomogeneities cancels in the differences in (\ref{10}). For a
spaceship travelling through the solar system, however, the
variation of the potential of the Sun and Earth must be included,
the second being negligible, but remember that the space change of
$\Phi$ does not produce any change in the frequency. Let us assume
now that all the matter (ordinary plus dark) and dark energy are
uniformly distributed. Since the distances between the galaxies
are increasing, $\Phi _{\rm av}(t)$ becomes less negative (or more
positive) as the time goes on, a consequence being that the
quantum vacuum thins down, its density being a decreasing function
of time. Hence the decrease of the permittivity and permeability
of empty space and the increase of the light velocity. The
consequences of these ideas agree well with the observations (see
fig. 1 in \cite{Ran02}).

 It must be stressed
that, as was argued in \cite{Ran02}, eqs. (\ref{10})-(\ref{20})
are not {\em ad hoc} assumptions but unavoidable consequences of
the fourth Heisenberg relation. We can average eq. (\ref{10}),
writing instead
\begin{equation}
\epsilon _{\rm r}=1-\beta [\Phi_{\rm av}(t) -\Phi _{\rm av}
(t_0)]/c^2,\;\;\;\; \mu _{\rm r}=1-\gamma [\Phi_{\rm av}(t) -\Phi
_{\rm av}(t_0)]/c^2, \label{20}
\end{equation}
where $\Phi _{\rm av}(t)$ is the space averaged gravitational
potential of all the universe at time $t$ and $t_0$ is the present
time ({\em i.e.} the age of the universe).

 Let $\Phi
_0$ is the gravitational potential produced by the critical
density distributed up to the distance of $R_U$ ($\Phi _0=-\int
_0^{R_U} G\rho _{\rm cr}4\pi r{\rm d}r\simeq -0.3c^2$ if
$R_U\approx 3,000$ Mpc)
  and let $\Omega _M,\,\Omega _\Lambda$ be the corresponding
present time relative densities of matter (ordinary plus dark) and
dark energy corresponding to the cosmological constant $\Lambda$.
Because of the expansion of the universe, the gravitational
potentials due to matter and dark energy equivalent to the
cosmological constant vary in time as the inverse of the scale
factor $a(t)$ and as its square $a^2(t)$, respectively. It turns
out therefore that
\begin{equation}
\Phi_{\rm av}(t)-\Phi_{\rm av}(t_0)=\Phi _0F(t),\;\;\;\mbox{ with }\;\;
  F(t)=\Omega
_M[1/a(t)-1]-2\Omega_\Lambda [a^2(t)-1],
\label{30}
\end{equation}
where $F(t_0)=0$, $\dot{F}(t_0)= -(1+3\Omega _\Lambda )H_0$. Let
us assume a universe with flat sections $t=\mbox{constant}$ ({\it
i.e.} $k=0$),  with $\Omega _M=0.3$, $\Omega _\Lambda =0.7$ and
Hubble parameter to $H_0=71\mbox{ km}\cdot \mbox{s}^{-1}\cdot
\mbox{Mpc}^{-1}= 2.3\times 10^{-18}\mbox{ s}^{-1}$. To find the
evolution of the average quantities, $\Phi_{\rm av}(t)-\Phi_{\rm
av}(t_0)$ must be substituted for $\Phi ({\rm r},t) -\Phi _{\rm
E}$ in (\ref{15}), what gives for the time evolutions of the fine
structure constant and the light velocity, at first order in the
potential,
\begin{equation}
\alpha (t)= \alpha \left[1+(3\beta -\gamma)F(t) \Phi _0/(2
c^2)\right], \;\;\;\; c(t)= c\left[1+(\beta +\gamma )F(t)\Phi
_0/(2 c^2)\right], \label{60}
\end{equation}
$c(t),\, \alpha (t)$ being the time evolutions and $\alpha =\alpha
(t_0),\, c=c(t_0)$, the present time values, {\em i.e.} the
constants that appear in the tables. It was shown in ref.
\cite{Ran02} that the first eq. (\ref{60}) gives a good agreement
with the observations by Webb {\em et al}  if $\xi =(3\beta
-\gamma )/2= 1.3\times 10^{-5}$ if $(\Omega _M,\, \Omega _\Lambda)
=(0.3,\, 0.7)$ (resp. $(1,0)$) (see fig. 1 in \cite{Ran02}).

{\bf The adiabatic acceleration of light.} It follows from the second eq.
(\ref{60}) that the  velocity of light increases in time, the
present value of the acceleration $a=\dot{c}(t_0)$ being equal to
\begin{equation}
a= -H_0c(\beta +\gamma )(1+3\Omega _\Lambda ) \Phi _0/(2c^2).
\label{80}
\end{equation}
Note that, as $H_0c= 6.9\times 10^{-10}\mbox{ m/s}^2$, $a$ is of
the  same order as the Pioneer acceleration $a_{\rm P}$ if $\beta$
and $\gamma$ are close to 1 and $\Omega _\Lambda =0.7$. It was
shown in \cite{Ran02} that the observed cosmological variation of
$\alpha$ can be explained with a value for $\xi =(3\beta -\gamma
)/2$ of the order of $10^{-5}$. We will show now that this same
model suggests an explanation of the anomalous Pioneer
acceleration as an effect of the quantum vacuum if $(\beta +\gamma
)/2$ has a
 value of the order of one, what we assume as a working hypothesis to be the case.

{\bf The adiabatic acceleration of light implies a blue shift.} It
will be shown now that the frequency $\omega _0$ of a
monochromatic light wave with such an adiabatic acceleration $a$
increases so that its time derivative $\dot{\omega}$ satisfies
\begin{equation}
\dot{\omega}/ \omega_0 =a/c. \label{90}
\end{equation}
Furthermore, an adiabatic acceleration of light has the same radio signature as a blue shift of the emitter, although a peculiar blue shift with no change of the wavelength ({\em i.e.} all the increase in velocity is used to increase the frequency).

 Equations (\ref{20})-(\ref{30}) tell that the time derivatives of the permittivity
 $\epsilon =\epsilon _{\rm r}\epsilon _0$ and permeability $\mu =\mu _{\rm r}\mu _0$ of empty space at present time $t_0$ are equal to
\begin{equation}
\dot{\epsilon} = \epsilon _0\beta (1+3\Omega _\Lambda )H_0 (\Phi
_0/c^2),\;\;\; \dot{\mu} = \mu _0\gamma (1+3\Omega _\Lambda )H_0
(\Phi _0/c^2). \label{100}
\end{equation}
These two derivatives are negative and very small.  To study the
propagation of the light in a medium whose permittivity and
permeability decrease adiabatically, we must take the Maxwell
equations and deduce the wave equations for the electric field
${\bf E}$ and the magnetic intensity $\bf H$, which are $\nabla
^2{\bf E}-{\partial _t}\left( \mu {\partial _t}(\epsilon {\bf
E})\right)=0$, $\nabla ^2{\bf H}-{\partial _t}\left( \epsilon
{\partial _t}(\mu {\bf H})\right)=0,$ or, more explicitly,
\begin{eqnarray}
\nabla ^2{\bf E}-{\partial _t^2}{\bf E}/c^{2}(t) &-&
\left({\dot{\mu}/ \mu _0}+{2\dot{\epsilon}/\epsilon
_0}\right){\partial _t{\bf E}}/c^2(t)
-{\dot{\epsilon}\dot{\mu}}{\bf E}/(\epsilon _0\mu _0c^2(t))=0,\label{120}\\
\nabla ^2{\bf H}-{\partial _t^2{\bf H}}/ c^2(t) &-&
\left({2\dot{\mu}/\mu _0}+{\dot{\epsilon}/\epsilon
_0}\right){\partial _t{\bf H}/ c^2(t)}
-{\dot{\epsilon}\dot{\mu}{\bf H}/( \epsilon _0\mu _0 c^2(t))}
=0,\label{130}
\end{eqnarray}
since at present time $\epsilon _{\rm r}=1,\;\mu _{\rm r}=1$.
Because of (\ref{100}), $\dot{\epsilon}/\epsilon _0$ and
$\dot{\mu}/\mu _0$ are of order $H_0=2.3\times 10^{-18}\mbox{
s}^{-1}$, so that the third and the fourth terms in the LHS of
(\ref{120}) and (\ref{130}) can be neglected for frequencies
$\omega \gg H_0$, in other words for any practical purpose.  We
are left with two classical wave equations with time dependent
light velocity $c(t)$.
\begin{equation}
\nabla ^2{\bf E}-{\partial _t^2{\bf E}/ c^2(t)}=0,\;\; \nabla
^2{\bf H}-{\partial _t^2{\bf H}/ c^2(t)}=0. \label{140}
\end{equation}
In order to find the behavior of a monochromatic light beam
according to these two wave equations, we start with the first one
and take ${\bf E}={\bf E}_0\exp[{-i(\kappa z-(\omega
_0+\dot{\omega}t/2)t)}]$, where the frequency is the time
derivative of the phase of {\bf E}, {\em i.e.} $\omega
_0+\dot{\omega}t$. Neglecting the second time derivatives and
working at first order in $\dot{\omega}$ (with $\dot{\omega}t\ll
\omega_0$, $\dot{\omega}\ll \omega _0^2$), substitution in
(\ref{120}) gives $\kappa ^2 =\left[(\omega_0
+\dot{\omega}t)^2-i\dot{\omega}\right]/c^2(t)$. It follows that
$\kappa =k+i \zeta = \pm (\omega_0/
c(t))[1+4\dot{\omega}t/\omega_0](\cos \varphi + i \sin\varphi ),$
with $\varphi =-\dot{\omega}/2\omega_0 ^2,$ so that $k=\pm
(\omega_0 /c)\, (1+\dot{\omega}t/ \omega_0)/(1+at/ c)$ what
implies $k=\pm \omega _0/c$ and eq. (\ref{90}), $\dot{\omega}
/\omega_0= a/c$,  as stated before. Also, $\zeta
=-\dot{\omega}/2\omega_0 c =a/2c^2$. The wave amplitude decreases
in the direction of propagation as $e^{-z/\ell}$ with
$\ell=2c^2/a$, but as $a$ is of order $H_0c$, $\ell$ is of order
of 5,000 Mpc, so that this attenuation can be neglected. It is
easy to show that to take $k +\dot{k}t$ for the wave vector  leads
to $\dot{k}=0$. These results are valid both for the solutions of
(\ref{120}) and (\ref{130}).

This shows that the electromagnetic waves verify eq. (\ref{90}),
so that $k$, and  the wavelength $\lambda$ therefore, remain
constant while the frequency increases with the same relative rate
as the light velocity. Note an important point: in a measurement
of the frequency of radiowaves , a blue shift is found (unrelated
to the velocity of the emitter), but optical observations of the
wavelength fail to find any effect.

{\bf Non-mechanical and non-gravitational explanation of the
Pioneer acceleration.} In this model, the Pioneer effect is
neither gravitational nor mechanical (it is not produced by any
force) but electromagnetic. What Anderson {\em et al} observed was
a blue shift in the radiowaves from the two Pioneers. More
precisely, a drift of the Doppler residuals  corresponding to a
positive constant time derivative of the frequency received from
the spaceships such that
\begin{equation} {\dot{\omega}/\omega_0} ={a_{\rm P}/ c},
\label{160} \end{equation} with $a_{\rm P}\approx 8.5\times
10^{-10}\mbox{ m/s}$. Obviously, the simplest interpretation of
this observation is that there is an unexpected acceleration of
the ship towards the Sun, due to an extra force alien to Newton
law of Gravitation. However, as we have seen in the previous
section (eq. (\ref{90})), this could be also the signature of the
acceleration of light. Indeed if $(\beta +\gamma )/2$ is close to
1, then eq. (\ref{80}) implies that the acceleration of light is
close to  $H_0c\approx 6.9\times 10^{-10}\mbox{ m/s}^2=0.8\,a_{\rm
P}$ (in the case $\Omega _M=0.3,\, \Omega _\Lambda =0.7$).

This would explain why the effect is not seen in the planets.
Indeed, the cartography of the solar system, being based on radar
ranging methods that measure the delay of round trips of
electromagnetic waves, is quite independent of an eventual
acceleration of light equal to $a_{\rm P}$, too small to have any
detectable influence. To be more precise let us consider two radar
ranging observations in which the flight time of the light is
measured. If the second observation is made one year after the
first, the relative difference between the two results would be
about $ \mbox{ 1 year }\times a_{\rm P}/c \simeq O(10^{-10})$,
which has a completely negligible effect on the measurement. For
instance, the difference and the sum of the radii of the Mars and
Earth orbits are known since the radar ranging studies of the
Viking missions with precisions of 100 m and 150 m, respectively
\cite{And98}. If the same observations had been repeated a year
later, the changes of these lengths due to an acceleration $a_{\rm
P}$ of the light  would have been close to \mbox{$10^{-8}$ m} and
$1.5\times 10^{-8}$ m, respectively, quite unobservable. This
gives a simple explanation of the riddle that the Pioneer effect
is observed in the spaceships but not in the planets.

 {\bf Comparison with the experiments.} One
could fear at first sight that this effect could be in conflict with the various experimental
tests that put stringent bounds to any departure from special
relativity or the equivalence principle \cite{Wil93}, the most
important being here the E\"otv\"os, the
Hughes-Drever and the gravitational redshift experiments. Let us see which are
the bounds that they impose on $\beta$ and $\gamma$. It has
been acknowledged that a variation of $e$ could lead to a
violation of the equivalence principle, since a small part of the
mass of a body would change in a way that depends on its chemical
composition \cite{Uza02,Bek02}. Indeed, according to von Weizs\"acker
semiempirical mass formula, there is a Coulomb
contribution to the mass of a nucleus $m$ given by $m_ {\rm C}= 3e^2Z(Z-1)/20 \pi \epsilon _0
r_0 A^{1/3}$, with $r_0\simeq 1.5\mbox{ fm}$,
 plus the electromagnetic mass of each of the protons. The ratio  $u=m_{\rm C}/m$
is of the order $10^{-3}$ and increases with $Z$. In this model,
the difference of the values of the mass of a body at two points
would be therefore
 $\Delta m =\Delta m_{\rm C}= 2\beta um\Delta \Phi /c^2$, $u$ being
the average value of $m_{\rm C}/m$ of its nuclei, which depends on
its chemical composition. Note that $|\nabla \Phi |/c^2\simeq
10^{-16}$ at Earth, the field our planet being the dominant part.
In an E\"otv\"os experiment, the contribution of the effect
proposed in this model to the E\"otv\"os ratio $\eta$
 would satisfy
$\eta \lesssim 2\beta u|\nabla  \Phi \cdot {\bf h}|/c^2$, where
{\bf h} is a vector between the two positions of the balance (the
time variation of $\Phi$ can be neglected here). Assuming that
$h<1$ m one has $\eta \lesssim \beta \times 10^{-18}$, while the
best bound in this type of experiments is $\eta < 10^{-12}$
(obtained by Roll, Krotkov and Dicke and Braginski), from which
$\beta < 10^{+6}$. No problem for this model.

The Hughes-Drever experiments \cite{Hug60,Dre61} were devised as
tests of the Mach principle.  By observing the Zeeman effect in
nuclei, they establish the bound $\Delta m/m <10^{-23}$, $\Delta
m$ being the anisotropic part of the mass of a nucleus. Although
the mass is technically isotropic in this model (it is always a
scalar), a certain anisotropy arises in the sense that the
electromagnetic mass of an nucleus changes differently along the
diverse directions around a point. The above given expression for
the electromagnetic mass must be used then, assuming a
displacement of the order of the diameter of a nucleus, and taking
the potential of the Earth (again the main contribution). One
finds thus easily that the effect proposed in this model gives a
contribution $\Delta m/m\lesssim \beta \times 10^{-32}$  to the
difference of the relative changes of the electromagnetic mass of
a nucleus between two directions in a terrestrial laboratory. The
corresponding restriction for the model is $\beta < 10^{+9}$. No
problem again.

The gravitational redshift of the frequency is given as $\Delta
\omega /\omega = -(1+\delta )\Delta \Phi /c^2$ with $\delta =0$ (a
non-zero value would indicate a violation of the equivalence
principle). Several experimental tests set bounds for $\delta$,
the best being $|\delta|\lesssim 2 \times 10^{-4}$ \cite{Wil93}.
It was obtained by Vessot and Levine \cite{Ves79,Ves80} with the
1420 MHz line of the hyperfine spectrum of Hydrogen ({\it i.e.}
measuring frequencies), between a terrestrial laboratory and a
rocket travelling upwards until a height of 10,000 km. The
frequency of that line here at Earth is $\omega _{\rm E}=8\alpha
^4g_p m^2c^2/3M\hbar$, $m$ and $M$ being the electron and proton
masses and $g_{\rm p}=2.79$ \cite{Sak67}. Assuming that the rest
energy of the electron is of electromagnetic origin \cite{Sch66},
then $m\propto e^2/c^2$, so that $\Delta m/m = (\beta -\gamma
)\Delta \Phi /c^2$, and $\omega _{\rm E}\propto e^4\alpha ^4/c^2$.
Hence the line would be emitted at spacetime point $({\bf r},t)$
with the frequency $\omega ({\rm r},t) = \omega _{\rm E}[1+4\Delta
e/e + 4\Delta \alpha /\alpha -2\Delta c/c]$, so that $\Delta
\omega /\omega =(\omega ({\rm r},t)-\omega _{\rm E})/\omega _{\rm
E} =[4\beta +4\xi-(\beta +\gamma)]\Delta \Phi /c^2=6\xi \Delta
\Phi /c^2$. Note that the effect of the acceleration $a$ is
negligible here, since the change of $\Phi$ during the short time
flight of the ray ($\leq 0.033$ s) is very small. In other words,
the line is produced with a slightly different frequency $\omega
({\bf r},t)$, and travels after until the receiver with constant
frequency, as in an optical medium. To be specific, if $\Phi ({\rm
r},t)<\Phi _{\rm E}$, then the line is seen at Earth with
frequency $\omega ({\rm r},t)\;(<\omega _{\rm E})$ and conversely.
The effect described in this model would produce, therefore, a
shift corresponding to $\delta =6\xi \simeq 8\times 10^{5}$, below
the Vessot-Levine bound (near borderline at most). Conclusion:
there is no conflict between this model and the experimental tests
of the special relativity and the equivalence principle.

{\bf Summary and conclusion.} Using a Newtonian approximation, the
model presented in reference \cite {Ran02} offers a unified
picture that accounts both for the cosmological variation of the
fine structure constant, observed by  Webb {\em et al}
\cite{Web01}, and for the anomalous Pioneer acceleration, observed
by Anderson {\em et al} \cite{And98}. More precisely, it explains
these two phenomena as due to the progressive attenuation of the
quantum vacuum because of the fourth Heisenberg relation combined
with its gravitational interaction with all the expanding
universe. In the first case, the fine structure constant increases
in time because a thinner vacuum implies a lesser renormalization
of the electron charge and an acceleration of light, the resulting
value of $\alpha$ being an increasing function of time. In the
second, because an acceleration of light has the same radio
signature as a blue shift of the frequency. In both cases the
effect depends on the coefficients $\beta$ and $\gamma$ in eqs.
(\ref{10}) which express the permittivity and the permeability of
the quantum vacuum as functions of the gravitational potential at
first order. It must be stressed that this model does not conflict
with the experimental tests of special relativity or the
equivalent principle. Indeed, Einstein second postulate of special
relativity would be still valid as an extremely good
approximation, its practical value being unaffected. The effect
was not observed in the planets  because they were not submitted
to the same kind of observation as the Pioneers.  The Pioneer
acceleration would be thus a manifestation of the universal
expansion (see \cite{Ros98} for another model with this in
common). The experimental test of this idea is surely difficult.
One way would be
 to repeat the measurements of $c$ during several years, another to measure the frequency
of the radiowaves emitted by a very stable source (not necessarily
in a spaceship) during a sufficiently long time.

Note that this model is free of {\em ad hoc} assumptions and does
predict, using well known basic laws of physics, (i) that the time
dependence of $\alpha$ is given by the function $F(t)$ in eqs.
(\ref{30}), what agrees with the observations (see fig. 1 in
\cite{Ran02}); (ii) that a blue shift must be seen in the radio
signal of any spaceship moving away from the Sun, quite similar to
the shift due to an extra acceleration of the ship towards the Sun
but unrelated to its motion.
 The change in the light velocity during one year if its acceleration is $a_{\rm P}$
would be just about 2.7 cm/s, only after 37 years the change would
amount to 1 m/s.

 To summarize, {\em this letter proposes as a possibility worth of consideration that the
 Pioneers did not suffer any extra acceleration but, quite on the
 contrary, that they followed the standard Newton laws, the observed and unmodelled
acceleration $a_{\rm P}$ being an observational effect of an
acceleration of light $a$ equal to $a_{\rm P}$, due to its
interaction with the quantum vacuum.}

I am grateful to Profs. F. Barbero, A.I. G\'omez de Castro, J.
Julve, M. Moles,  A. Tiemblo and J. L. Trueba for discussions.


\begin{thebibliography}{Ran92b}
\bibitem{Ran02}  RA\~NADA A. F., {\em Europhys. Lett.} {\bf 61}
(2003) 174.

\bibitem{Web01}
 WEBB J.K.,  MURPHY M.T.,  FLAMBAUM V.V.,  DZUBA V.A.,  BARROW
J.D.,  CHURCHILL C.W.,  PROCHASKA J.X. and  WOLFE A.M., {\em Phys.
Rev. Lett.} {\bf 87} (2001) 091301.

\bibitem{And98}  ANDERSON J.D.,  LAING PH. A.,  LAU E.L.,  LIU A.S.,  MARTIN NIETO M. and  TURYSHEV S.G.,
{\em Phys. Rev. Lett.} {\bf 81} (1998) 2858.

\bibitem{And02} ANDERSON J.D.,  LAING PH. A.,  LAU E.L.,  LIU A.S.,  MARTIN NIETO M. and  TURYSHEV S.G.,
 {\em Phys. Rev. D} {\bf 65} (2002) 082004.

\bibitem{Ein12} EINSTEIN, A. (1912). {\em Ann. Physik} {\bf 38},
1059; reprinted in {\it The collected papers of Albert Einstein}
(1996), English translation, vol 4,  130 (Princeton University
Press, Princeton).

\bibitem{Wil93}   WILL C. M. (1993). {\it Theory and experiment in
gravitational physics}, revised edition (Cambridge University
Press, Cambridge).
\bibitem{Uza02}  UZAN, J. PH. (2003). {\it Rev. Mod. Phys.} {\bf 75}, 403.
\bibitem{Bek02} BEKENSTEIN, J. D. (2002). {\it Phys. Rev.} {\bf D 66},
123514.
 \bibitem{Hug60} HUGHES, V.W., ROBINSON, H.G. and BELTRAN
 LOPEZ, V. (1960). {\it Phys. Rev. Lett.} {\bf 4} , 342.
 \bibitem{Dre61} DREVER, R. W. P. (1961). {\it Phil. Mag.}, {\bf 6},
 683.

 \bibitem{Ves79} VESSOT, R.F.C. and LEVINE, M. W. (1979). {\it
 Gen. Rel. and Grav.} {\bf 10}, 181.
 \bibitem{Ves80} VESSOT, R.F.C., {\it et al}. (1980). {\it Phys.
 Rev. Lett.} {\bf 45}, 2081.
\bibitem{Sak67} SAKURAI, J.J. (1967) {\it Advanced quantum
mechanics} (Addison-Wesley, Reading).
\bibitem{Sch66} SCHWEBER, S.S. (1966). {\it An introduction to
relativistic quantum field theory} (Harper $\&$ Row, New York).

\bibitem{Ros98}  ROSALES, J.L. and  SANCHEZ-GOMEZ, J.L., gr-qc/9810085 v3, 24 May 1999.





\end{thebibliography}
\end{document}